 \documentclass[authoryear,preprint,review,12pt]{elsarticle}

\usepackage{amssymb,amsmath}
\usepackage{latexsym}
\usepackage{amsmath}
\usepackage{graphicx}

\journal{Mechanics of Materials}

\begin{document}

\begin{frontmatter}



\title{Universal Theorems for Total Energy of the Dynamics of Linearly Elastic Heterogeneous Solids}


\author{Ankit Srivastava and Sia Nemat-Nasser}

\address{Department of Mechanical and Aerospace Engineering\\ 
University of California, San Diego\\
La Jolla, CA, 92093-0416 USA}

\begin{abstract}
In this paper we consider a sample of a linearly elastic heterogeneous composite in \emph{elastodynamic} equilibrium and present universal theorems which provide lower bounds for the total elastic strain energy plus the kinetic energy, and the total complementary elastic energy plus the kinetic energy. For a general heterogeneous sample which undergoes \emph{harmonic motion at a single frequency}, we show that, among all consistent boundary data which produce the same average strain, the uniform-stress boundary data render the total elastic strain energy plus the  kinetic energy an absolute minimum. We also show that, among all consistent boundary data which produce the same average momentum in the sample, the uniform velocity boundary data render the total complementary elastic energy plus the  kinetic energy an absolute minimum. We do not assume statistical homogeneity or material isotropy in our treatment, although they are not excluded. These universal theorems are the dynamic equivalent of the universal theorems already known for the static case (\cite{nemat1995universal}). It is envisaged that the bounds on the total energy presented in this paper will be used to formulate computable bounds on the overall dynamic properties of linearly elastic heterogeneous composites with arbitrary microstructures.

\end{abstract}


\end{frontmatter}


\section{Introduction}

One of the main objectives of micromechanics is to estimate the overall properties of a heterogeneous composite in terms of the architectural and material properties of its micro-constituents. For the static case, this amounts to defining effective properties which relate the domain averages of the stress and strain tensors over a suitably large sample called a \emph{Representative Volume Element} (RVE; See \cite{hill1963elastic,hashin1965elasticity,kröner1977bounds,willis1981avariational,willis1981bvariational}). The analysis is complicated by the following twin problems:

1. It is often difficult to identify a suitable RVE that characterizes the composite.

2. In general the average values of the stress and strain tensors depend upon the boundary conditions to which the RVE is subjected. Therefore, effective properties defined to relate these averages depend upon the boundary conditions under which the averages are calculated.

To deal with these difficulties, \emph{statistical homogeneity} is used which assumes that the overall properties of the composite can be represented by those of an RVE. Furthermore, it implies that the overall response of the RVE is almost the same for any boundary condition as long as the average stress and strain tensors are kept constant. By employing statistical homogeneity, strain and complementary energy functionals can be defined to estimate the overall properties of the heterogeneous composite \cite{eshelby1957determination,hashin1962some,hashin1962variational,hashin1965elasticity,walpolea1966bounds,walpoleb1966bounds,korringa1973theory}.
However, such estimates have been shown to be merely plausible bounds and not rigorous bounds on the effective properties \cite{willis1981avariational}. It was subsequently shown that exact inequalities for the elastic and the complementary energies stored in a finite body under arbitrary boundary conditions could be established \cite{nemat1995universal,nemat1999micromechanics}. These bounds on the energies were, in turn, used to provide improvable and computable bounds on the static effective properties of heterogeneous composites.

The recent interest in metamaterials \cite{smith2000composite,sheng2003locally,liu2005analytic,milton2007modifications} has necessitated a systematic homogenization procedure to estimate the effective dynamic properties of composites by using field averaging or ensemble averaging techniques \cite{smith2006homogenization,amirkhizi2008microstructurally,amirkhizi2008numerical,willis2009exact,nemat2011homogenization,willis2011effective}. It has been shown that in the homogenized representation the average stress is coupled with the average velocity and that the average momentum is coupled with the average strain (See also \cite{shuvalov2011effective}). Recently, a micromechanical method to calculate the effective properties of periodic elastic composites was proposed \cite{nematnasser2011overall,srivastava2011overall}. This method provides effective parameters which reduce to the effective parameters calculated from the ensemble averaging technique of Ref. \cite{milton2007modifications} or the field averaging technique of Ref. \cite{nematnasser2011overall} when the dispersion relation of the composite is enforced. 

In this paper we begin with a brief overview of the elastostatic problem, stating the universal energy theorems for this case. For the elastodynamic case, we present the general form of the effective constitutive relations and briefly discuss the properties of the effective parameters which arise in the averaged constitutive relations. Finally, we present universal theorems which are the dynamic analogues of the static universal theorems presented in Ref. \cite{nemat1999micromechanics}. These universal theorems bound the total elastic energy plus the  kinetic energy and the total complementary energy plus the  kinetic energy of a composite in an elastodynamic state and pave the way for subsequently establishing rigorous bounds on the effective dynamic properties of the composite.

\section{Elastostatic Universal Theorems}

Consider the static case of a general heterogeneous solid which consists of various elastic phases. There is no restriction on the number, geometry, material, or orientation of each constituting microphase, i.e., neither statistical homogeneity nor isotropy is assumed. Consider an arbitrary finite sample of volume $\Omega$ and boundary $\partial\Omega$. The field variables for the problem are the stress, $\boldsymbol{\sigma}\mathbf{(x)}$, and strain, $\boldsymbol{\epsilon}\mathbf{(x)}$, tensors and the displacement vector, $\mathbf{u(x)}$. The constitutive equation relates the stress to the strain through the stiffness, $\mathbf{C(x)}$, or the compliance, $\mathbf{D(x)=C^{-1}}$, tensors; in what follows, the \textbf{x} dependence is implicit,

\begin{equation}
\boldsymbol{\sigma}=\mathbf{C}:\boldsymbol{\varepsilon};\quad\boldsymbol{\varepsilon}=\mathbf{D}:\boldsymbol{\sigma}
\label{S-EEqn}
\end{equation}

Strain is related to displacement through the kinematic relation (field equation),

\begin{equation}
\boldsymbol{\varepsilon}=\frac{1}{2}\left[\boldsymbol{\nabla}\mathbf{u}+\boldsymbol{\nabla}^T\mathbf{u}\right]
\end{equation}

For elastostatically admissible boundary data, effective properties (effective stiffness and compliance tensors) are defined by relating the domain averages of the stress and the strain tensors,

\begin{equation}
\langle\boldsymbol{\sigma}\rangle_\Omega=\mathbf{C}^{eff}:\langle\boldsymbol{\varepsilon}\rangle_\Omega;\quad\langle\boldsymbol{\varepsilon}\rangle_\Omega=\mathbf{D}^{eff}:\langle\boldsymbol{\sigma}\rangle_\Omega
\end{equation}

where the domain average is defined by,

\begin{equation}
\langle\mathbf{Q}\rangle_\Omega=\frac{1}{\Omega}\int_{\Omega}\mathbf{Q}d\Omega
\nonumber
\end{equation}

Effective properties defined above, depend upon the boundary conditions under which the strain and stress fields are generated. It is, therefore, of interest to study if there exist special boundary conditions which bound the effective properties associated with any boundary data. It was shown in Refs. \cite{nemat1995universal,nemat1999micromechanics} that the elastic energy and complementary energy associated with the elastostatic system are bound by special boundary conditions and that \emph{this fact could be used to place strict bounds on the effective parameters}. To this end, the following two stress/strain states were defined:

\begin{itemize}
\item \emph{Weakly Kinematically Admissible Strain Fields}: Any compatible strain field with a prescribed average value.
\item \emph{Weakly Statically Admissible Stress Fields}: Any compatible stress field with a prescribed average value.
\end{itemize}

And the following two universal theorems were proved:

\begin{itemize}
\item \emph{Universal theorem for elastic strain energy}: Among all weakly kinematically admissible strain fields, the strain field produced by uniform boundary tractions renders the total strain energy an absolute minimum.
\item \emph{Universal theorem for complementary elastic energy}: Among all weakly statically admissible stress fields, the stress field produced by linear displacement (uniform strain) boundary data renders the total complementary energy an absolute minimum.
\end{itemize}

In the present paper we show that analogous universal theorems exist for the elastodynamic case. These universal theorems may subsequently be used to place strict bounds on the effective dynamic parameters.

\section{Elastodynamic Universal Theorems}

Now consider the dynamic case of a general heterogeneous solid which consists of various elastic phases. As for the elastostatic case, there is no restriction on the number, geometry, material, or orientation of each constituting microphase, i.e., neither statistical homogeneity nor isotropy is assumed. Consider an arbitrary finite sample of volume $\Omega$ and boundary $\partial\Omega$. The field variables are represented by the stress, $\boldsymbol{\sigma}\mathbf{(x)}$, and strain, $\boldsymbol{\epsilon}\mathbf{(x)}$, tensors and the vectors of momentum, $\mathbf{p(x)}$, and velocity, $\mathbf{\dot{u}(x)}$. Constitutive relations relating the stress to the strain are given by Eqs. (\ref{S-EEqn}).  The relation between the momentum and the velocity at every point in $\Omega$ is,

\begin{equation}
\displaystyle \mathbf{p}=\rho\mathbf{\dot{u}}\\
\label{EFieldEqn1}
\end{equation}

where $\rho\mathbf{(x)}$ is the density. The field equations relate the stress to the momentum and the strain to the velocity,

\begin{equation}
\begin{array}{l}
\displaystyle {\nabla}\cdot\boldsymbol{\sigma}=\dot{\mathbf{p}}\\
\displaystyle \dot{\boldsymbol{{\varepsilon}}}=({\nabla}\mathbf{\dot{u}}+{\nabla}^T\mathbf{\dot{u}})\\
\end{array} 
\label{EFieldEqn}
\end{equation}

Dynamic homogenization is an active area of research (See \cite{norris1992dispersive,norris1992shear,norris1993waves,wang2002floquet,andrianov2008higher} and references therein.) Effective dynamic parameters are defined by relating the domain averages of the field variables. The general form of the averaged constitutive relation is given by (\cite{milton2007modifications,willis2009exact,nematnasser2011overall,willis2011effective,srivastava2011overall}),

\begin{equation}\label{DynamicRelation2}
\begin{array}{l}
\displaystyle \langle\boldsymbol{\varepsilon}\rangle_\Omega=\mathbf{\bar{D}}:\langle\boldsymbol{\sigma}\rangle_\Omega+\mathbf{S}^1\cdot\langle\mathbf{\dot{u}}\rangle_\Omega\\
\displaystyle \langle\mathbf{p}\rangle_\Omega=\mathbf{S}^2:\langle\boldsymbol{\sigma}\rangle_\Omega+\boldsymbol{\bar{\rho}}\cdot\langle\mathbf{\dot{u}}\rangle_\Omega
\end{array} 
\end{equation}

All the effective constitutive parameters are non-local in space and time. The effective parameters may be complex even if there is no dissipation in the system, the imaginary parts resulting from the asymmetries of, e. g., the unit cell of a periodic composite. $\mathbf{\bar{D}}$ is the fourth-order effective compliance tensor which has minor symmetries, $\bar{D}_{ijkl}=\bar{D}_{jikl}=\bar{D}_{ijlk}$. It does not have the major symmetry associated with the elasticity or the compliance tensor but has a hermitian relationship over the major transformation, $\bar{D}_{ijkl}=[\bar{D}_{klij}]^*$, where * denotes complex conjugation. Effective density is a second-order tensor with a hermitian relationship over the transformation of its indices, $\bar{\rho}_{ij}=[\bar{\rho}_{ji}]^*$, and $\mathbf{S}^1$, $\mathbf{S}^2$ are third-order coupling tensors with the relationship $S_{ijk}^1=[S_{kij}^2]^*$.

Defined as above, the effective parameters depend upon the boundary conditions under which the domain averages of the field variables have been calculated. It is, therefore, of interest to investigate the existence of special boundary conditions which may bound the effective parameters by bounding the associated total elastic energy plus the  kinetic energy and the total complementary elastic energy plus the  kinetic energy for any boundary data. We present two theorems which are the elastodynamic equivalent of the elastostatic universal theorems stated above. These theorems prove the existence of special boundary conditions under which the total elastic energy plus the  kinetic energy and the total complementary energy plus the  kinetic energy achieve their absolute minima. 

\subsection{Universal Theorems}

We define the following concepts of weak admissibility,

\begin{itemize}
\item \emph{Weakly Kinematically Admissible Strain Fields}: Any compatible strain field with a prescribed average value.
\item \emph{Weakly Dynamically Admissible Momentum Fields}: Any compatible momentum field with a prescribed average value.
\end{itemize}

It is not required for these fields to satisfy any specific boundary data, only their averages are required to be equal to specified values.

\subsection{Universal Theorem for Total Elastic Strain and Kinetic Energy}

Consider a weakly admissible strain field $\boldsymbol{\epsilon}(\mathbf{x})$ which satisfies $\langle\boldsymbol{\epsilon}\rangle_\Omega=\boldsymbol{\epsilon}^0$. The total elastic strain and kinetic energy associated with the body is given by,

\begin{equation}
\Pi(\boldsymbol{\epsilon};\boldsymbol{\epsilon}^0)=\frac{1}{2}\langle\boldsymbol{\epsilon}:\mathbf{C}:\boldsymbol{\epsilon}\rangle_{\Omega}+\frac{1}{2}\langle \mathbf{v}\cdot\rho \mathbf{v}\rangle_\Omega
\end{equation}

where the inner product is given by,

\begin{equation}
\begin{array}{l}
\displaystyle \langle\boldsymbol{\epsilon}:\mathbf{C}:\boldsymbol{\epsilon}\rangle_{\Omega}=\frac{1}{\Omega}\int_{\Omega}\boldsymbol{\epsilon}:\mathbf{C}:\boldsymbol{\epsilon}^*d\Omega\\

\displaystyle \langle\mathbf{v}\cdot\rho\mathbf{v}\rangle_{\Omega}=\frac{1}{\Omega}\int_{\Omega}\mathbf{v}\cdot\rho\mathbf{v}^*d\Omega\\

\end{array} 
\end{equation}

For any other weakly kinematically admissible strain field, $\hat{\boldsymbol{\epsilon}}(\mathbf{x});\langle\hat{\boldsymbol{\epsilon}}(\mathbf{x})\rangle_{\Omega}=\boldsymbol{\epsilon}^0$, the total energy is given by,

\begin{equation}
\Pi(\hat{\boldsymbol{\epsilon}};\boldsymbol{\epsilon}^0)=\frac{1}{2}\langle\hat{\boldsymbol{\epsilon}}:\mathbf{C}:\hat{\boldsymbol{\epsilon}}\rangle_{\Omega}+\frac{1}{2}\langle \hat{\mathbf{v}}\cdot\rho \hat{\mathbf{v}}\rangle_\Omega
\end{equation}

where $\hat{\mathbf{v}}\mathbf{(x)}$ is the velocity field resulting from the displacement field $\hat{\mathbf{u}}\mathbf{(x)}$ which corresponds to the strain $\hat{\boldsymbol{\epsilon}}(\mathbf{x})$. Since $\mathbf{C}$ has major symmetry, the difference in the total  energy for the two weakly admissible strain fields is,

\begin{equation}\label{Energy}
\begin{array}{l}
\displaystyle \Pi(\boldsymbol{\epsilon};\boldsymbol{\epsilon}^0)-\Pi(\hat{\boldsymbol{\epsilon}};\boldsymbol{\epsilon}^0)=\frac{1}{4}\langle(\boldsymbol{\epsilon}-\hat{\boldsymbol{\epsilon}}):\mathbf{C}:(\boldsymbol{\epsilon}-\hat{\boldsymbol{\epsilon}})\rangle_{\Omega} + \frac{1}{2}\langle(\boldsymbol{\epsilon}-\hat{\boldsymbol{\epsilon}}):\mathbf{C}:\hat{\boldsymbol{\epsilon}}\rangle_{\Omega}\\
\displaystyle \quad\quad\quad\quad\quad\quad\quad\quad+\frac{1}{4}\langle(\mathbf{v}-\hat{\mathbf{v}})\cdot\rho(\mathbf{v}-\hat{\mathbf{v}})\rangle_{\Omega} + \frac{1}{2}\langle(\mathbf{v}-\hat{\mathbf{v}})\cdot\rho\hat{\mathbf{v}}\rangle_{\Omega}+c.c.
\end{array} 
\end{equation}

where $c.c.$ denotes the complex conjugate of the preceding expression. Setting $\hat{\boldsymbol{\sigma}}=\mathbf{C}:\hat{\boldsymbol{\epsilon}}$ and $\tilde{\boldsymbol{\epsilon}}=(\boldsymbol{\epsilon}-\hat{\boldsymbol{\epsilon}})$, the second term on the right hand side can be written as,

\begin{equation}\nonumber
\frac{1}{2}\langle(\boldsymbol{\epsilon}-\hat{\boldsymbol{\epsilon}}):\mathbf{C}:\hat{\boldsymbol{\epsilon}}\rangle_{\Omega}=\frac{1}{2\Omega}\int_\Omega \tilde{\epsilon}_{ij}\hat{\sigma}_{ij}^*d\Omega
\end{equation}

Now we have,

\begin{equation}
\begin{array}{l}
\displaystyle \tilde{\epsilon}_{ij}\hat{\sigma}^*_{ij}=\frac{1}{2}(\tilde{u}_{i,j}+\tilde{u}_{j,i})\hat{\sigma}^*_{ij}=\frac{1}{2}(\tilde{u}_{i}\delta_{kj}+\tilde{u}_{j}\delta_{ki})_{,k}\hat{\sigma}^*_{ij}\\
\quad\quad\quad\quad=\frac{1}{2}\left[(\tilde{u}_{i}\delta_{kj}+\tilde{u}_{j}\delta_{ki})\hat{\sigma}^*_{ij}\right]_{,k}-\frac{1}{2}(\tilde{u}_{i}\delta_{kj}+\tilde{u}_{j}\delta_{ki})\hat{\sigma}^*_{ij,k}\\
\quad\quad\quad\quad=(\hat{\sigma}^*_{ij}\tilde{u}_j)_{,i}+\hat{\sigma}^*_{ij,j}\tilde{u_i}
\end{array} 
\end{equation}

For the static case the last term in the above equation is equal to zero based on the conservation law $\hat{\sigma}^*_{ij,j}=0$. For the dynamic case, however, we have $\hat{\sigma}^*_{ij,j}=\hat{\dot{p}}^*_i$, hence,

\begin{equation}\label{Decompose}
\begin{array}{l}
\displaystyle \frac{1}{2}\langle(\boldsymbol{\epsilon}-\hat{\boldsymbol{\epsilon}}):\mathbf{C}:\hat{\boldsymbol{\epsilon}}\rangle_{\Omega}=\frac{1}{2\Omega}\int_\Omega \nabla\cdot({\hat{\boldsymbol{\sigma}}}^*\cdot\tilde{\mathbf{u}})d\Omega-\frac{1}{2\Omega}\int_\Omega \hat{\dot{\mathbf{p}}}^*\cdot\tilde{\mathbf{u}}d\Omega\\
\displaystyle \quad\quad\quad\quad\quad\quad\quad\quad=\frac{1}{2\Omega}\int_{\partial\Omega}\mathbf{n}\cdot\hat{\boldsymbol{\sigma}}^*\cdot\tilde{\mathbf{u}}d\partial\Omega-\frac{1}{2\Omega}\int_\Omega \hat{\dot{\mathbf{p}}}^*\cdot\tilde{\mathbf{u}}d\Omega
\end{array} 
\end{equation}

Now the last term in Eq. (\ref{Energy}) is, 

\begin{equation}
\frac{1}{2}\langle(\mathbf{v}-\hat{\mathbf{v}})\cdot\rho\hat{\mathbf{v}}\rangle_{\Omega}=\frac{1}{2\Omega}\int_\Omega \tilde{\dot{\mathbf{u}}}\cdot\hat{\mathbf{p}}^*d\Omega
\end{equation}

This cancels with the volume integral in Eq. (\ref{Decompose}) if the two cases for which the energies are being compared are in harmonic motion with the same frequency. Moreover if the boundary conditions associated with the strain field $\hat{\boldsymbol{\epsilon}}$ are such that, on the boundary,


\begin{equation}
\mathbf{n(x)}\cdot\hat{\boldsymbol{\sigma}}\mathbf{(x)}|_{\partial\Omega}=\mathbf{n(x)}\cdot{\boldsymbol{\Sigma}}
\end{equation}

where ${\boldsymbol{\Sigma}}$ is a constant tensor, then the surface integral in Eq. (\ref{Decompose}) can be written as,

\begin{equation}
\frac{1}{\Omega}\int_{\partial\Omega}\mathbf{n}\cdot\hat{\boldsymbol{\sigma}}^*\cdot\tilde{\mathbf{u}}d\partial\Omega=\langle(\boldsymbol{\epsilon}-\hat{\boldsymbol{\epsilon}})\rangle_\Omega:\boldsymbol{\Sigma}^*
\end{equation}

This is zero when both $\boldsymbol{\epsilon}$ and $\hat{\boldsymbol{\epsilon}}$ are weakly kinematically admissible fields with volume averages equal to $\boldsymbol{\epsilon}^0$. Similar considerations apply to the complex conjugate parts of Eq. (\ref{Energy}) and it can be shown that the remaining terms in Eq. (\ref{Energy}) are always real and positive given the symmetric, positive-definiteness of $\mathbf{C}$ and the scalar nature of $\rho$.

The above treatment shows that among all weakly kinematically admissible strain fields, the sum of the elastic strain and the kinetic energy contained in a finite body, \emph{in harmonic motion at a common frequency}, is minimum for the case when the boundary conditions are one of uniform traction (constant stress). To summarize, our first universal theorem for the total energy is,

\emph{In elastodynamics, among all weakly kinematically admissible strain fields at a given frequency, the strain field produced by uniform boundary tractions renders the total strain energy plus the kinetic energy an absolute minimum.}

\subsection{Universal Theorem for Total Complementary Energy}

Now consider a weakly admissible momentum field ${{\mathbf{p}}}(\mathbf{x})$ which satisfies $\langle\mathbf{p}\rangle_\Omega=\mathbf{p}^0$. The total complementary elastic energy plus the kinetic energy associated with the body is given by,


\begin{equation}
\Pi^c(\mathbf{p};\mathbf{p}^0)=\frac{1}{2}\langle\boldsymbol{\sigma}:\mathbf{D}:\boldsymbol{\sigma}\rangle_{\Omega}+\frac{1}{2}\langle \mathbf{v}\cdot\rho \mathbf{v}\rangle_\Omega
\end{equation}

where $\mathbf{D(x)}=\mathbf{[C(x)]}^{-1}$ is the compliance. For any other weakly admissible momentum field, $\hat{\mathbf{p}}(\mathbf{x});\langle\hat{\mathbf{p}}(\mathbf{x})\rangle_{\Omega}=\mathbf{p}^0$, the total energy is now given by,

\begin{equation}
\Pi^c(\hat{\mathbf{p}};\mathbf{p}^0)=\frac{1}{2}\langle\hat{\boldsymbol{\sigma}}:\mathbf{D}:\hat{\boldsymbol{\sigma}}\rangle_{\Omega}+\frac{1}{2}\langle \hat{\mathbf{v}}\cdot\rho \hat{\mathbf{v}}\rangle_\Omega
\end{equation}

The difference in the total complementary energy of these states is,

\begin{equation}\label{EnergyC}
\begin{array}{l}
\displaystyle \Pi^c(\mathbf{p};\mathbf{p}^0)-\Pi^c(\hat{\mathbf{p}};\mathbf{p}^0)=\frac{1}{4}\langle(\boldsymbol{\sigma}-\hat{\boldsymbol{\sigma}}):\mathbf{D}:(\boldsymbol{\sigma}-\hat{\boldsymbol{\sigma}})\rangle_{\Omega} + \frac{1}{2}\langle(\boldsymbol{\sigma}-\hat{\boldsymbol{\sigma}}):\mathbf{D}:\hat{\boldsymbol{\sigma}}\rangle_{\Omega}\\
\displaystyle \quad\quad\quad\quad\quad\quad\quad\quad+\frac{1}{4}\langle(\mathbf{v}-\hat{\mathbf{v}})\cdot\rho(\mathbf{v}-\hat{\mathbf{v}})\rangle_{\Omega} + \frac{1}{2}\langle(\mathbf{v}-\hat{\mathbf{v}})\cdot\rho\hat{\mathbf{v}}\rangle_{\Omega}+c.c.
\end{array} 
\end{equation}

Denoting $\hat{\boldsymbol{\epsilon}}=\mathbf{C}:\hat{\boldsymbol{\sigma}}$ and $\tilde{\boldsymbol{\sigma}}=(\boldsymbol{\sigma}-\hat{\boldsymbol{\sigma}})$, the second term on the right hand side can be written as,

\begin{equation}\label{sub2vol}
\frac{1}{2}\langle(\boldsymbol{\sigma}-\hat{\boldsymbol{\sigma}}):\mathbf{D}:\hat{\boldsymbol{\sigma}}\rangle_{\Omega}=\frac{1}{2\Omega}\int_\Omega \tilde{\sigma}_{ij}\hat{\epsilon}^*_{ij}d\Omega
\end{equation}

As shown in the previous subsection, we have

\begin{equation}
\tilde{\sigma}_{ij}\hat{\epsilon}^*_{ij}=(\tilde{\sigma}_{ij}\hat{u}^*_j)_{,i}+\tilde{\sigma}_{ij,j}\hat{u}^*_i
\end{equation}

so that Eq. (\ref{sub2vol}) becomes,

\begin{equation}\label{sub2Decompose}
\begin{array}{l}
\displaystyle \frac{1}{2}\langle(\boldsymbol{\sigma}-\hat{\boldsymbol{\sigma}}):\mathbf{D}:\hat{\boldsymbol{\sigma}}\rangle_{\Omega}=\frac{1}{2\Omega}\int_\Omega \nabla\cdot({\tilde{\boldsymbol{\sigma}}}\cdot\hat{\mathbf{u}}^*)d\Omega-\frac{1}{2\Omega}\int_\Omega \tilde{\dot{\mathbf{p}}}\cdot\hat{\mathbf{u}}^*d\Omega\\
\displaystyle \quad\quad\quad\quad\quad\quad\quad\quad=\frac{1}{2\Omega}\int_{\partial\Omega}\mathbf{n}\cdot\tilde{\boldsymbol{\sigma}}\cdot\hat{\mathbf{u}}^*d\partial\Omega-\frac{1}{2\Omega}\int_\Omega \tilde{\dot{\mathbf{p}}}\cdot\hat{\mathbf{u}}^*d\Omega
\end{array} 
\end{equation}

The volume integral in the above equation cancels the last term on the right side of Eq. (\ref{EnergyC}) if the two weakly admissible cases are harmonic motion with the same frequency. Therefore, for harmonic motion of the same frequency, say, $\omega$, if the boundary condition for the second case is such that

\begin{equation}
\mathbf{\hat{\dot{u}}(x)}|_{\partial\Omega}=\mathbf{\dot{U}}=-i\omega\mathbf{{U}}
\end{equation}

then the surface integral in Eq. (\ref{sub2Decompose}) can be written as,

\begin{equation}
\frac{1}{\Omega}\int_{\partial\Omega}\mathbf{n}\cdot\tilde{\boldsymbol{\sigma}}\cdot\hat{\mathbf{u}}^*d\partial\Omega=-\frac{\dot{U}^*_i}{i\omega\Omega}\int_{\Omega}\tilde{\sigma}_{ij,j}d\Omega=-\frac{\dot{U}^*_i}{i\omega\Omega}\int_{\Omega}\tilde{\dot{p}}_id\Omega
\end{equation}

Invoking the harmonic nature of the motion, we can write the final term in the above equations as,

\begin{equation}
-\frac{\dot{U}^*_i}{i\omega\Omega}\int_{\Omega}\tilde{\dot{p}}_id\Omega=\frac{\dot{U}^*_i}{\Omega}\int_{\Omega}\tilde{{p}}_id\Omega
\end{equation}

The above integral goes to zero when the momentum fields for the two cases have the same average value. Similar comments apply to the complex conjugate part of Eq. (\ref{EnergyC}). In light of this, our second universal theorem is,

\emph{In elastodynamics, among all weakly dynamically admissible momentum fields at a given frequency, the momentum field produced by uniform boundary velocities renders the total complementary energy plus the kinetic energy an absolute minimum.}

\section{Conclusions}

The static energy theorems given in Ref. \cite{nemat1999micromechanics} proved that the elastic energy and the complementary energy of a heterogeneous solid corresponding to any boundary conditions, are bounded from below by that corresponding to special boundary conditions, i.e., constant stress and linear displacement, respectively. Thus, the energy theorems show that the effective properties are bounded by the effective properties defined for these special boundary data. This was then used to calculate strict and computable bounds on the static effective parameters. For the dynamic case we, have proved the existence of analogous theorems which bound the total strain energy plus the kinetic energy, and the total complementary energy plus the kinetic energy. It is worth to note that the boundary condition which provides a bound for the strain energy in the static case, i.e., the constant boundary tractions, also provides a bound for the total strain energy plus the kinetic energy for the dynamic case. On the other hand, the boundary condition which provides a bound for the complementary energy in the static case, i. e., linear displacements (or constant strain) does not provide a bound for the total complementary energy plus the kinetic energy in the dynamic case. Instead now, for a common average momentum,  it is the uniform velocity boundary data that provide the bound for the total complementary energy plus the kinetic energy of the elastic composite. The energy bounds proved for the elastodynamic case in this paper are expected to provide strict bounds on the effective dynamic properties of heterogeneous composites.

\section{Acknowledgement}

This research has been conducted at the Center of Excellence for Advanced Materials (CEAM) at the University of California, San Diego, under DARPA AFOSR Grants FA9550-09-1-0709 and RDECOM W91CRB-10-1-0006 to the University of California, San Diego.

\section{References}







\end{document}